\documentclass[final,12pt,5p,twocolumn]{elsarticle}
\usepackage[fleqn]{amsmath}
\usepackage{amssymb}
\usepackage{amsthm}
\usepackage{graphics}
\usepackage{graphicx}
\usepackage{lineno}
\usepackage{float}
\usepackage{mathrsfs}
\usepackage{parskip}
\usepackage{lipsum}
\usepackage{mathtools}
\usepackage{cuted}

\makeatletter
\newcommand{\mathleft}{\@fleqntrue\@mathmargin\parindent}
\newcommand{\mathcenter}{\@fleqnfalse}
\makeatother

\journal{Journal of Computational Science}

\begin{document}

\begin{frontmatter}

\title{Explosive Synchronization in Multilayer Dynamically Dissimilar Networks\tnoteref{label0}}
%\tnotetext[label0]{}

\author[label1,label3]{Sarika Jalan\corref{cor1}}
\cortext[cor1]{sarikajalan9@gmail.com}
%\ead[url]{http://www.iiti.ac.in/people/~sarika/}

\author[label1]{Ajay Deep Kachhvah}
%\ead{author.two@mail.com}

\address[label1]{Complex Systems Lab, Discipline of Physics, Indian Institute of Technology Indore, Khandwa Road, Simrol, Indore-453552, India}
\address[label3]{Center for Theoretical Physics of Complex Systems, Institute for Basic Science (IBS), Daejeon 34126, Republic of Korea}

\author[label4]{Hawoong Jeong}
\address[label4]{Department of Physics, Center for Complex Systems, Korea Advanced Institute of Science and Technology, Daejeon 34141, Republic of Korea}

\begin{abstract}
The phenomenon of explosive synchronization, which originates from hypersensitivity to small perturbation caused by some form of frustration prevailed in various physical and biological systems, has been shown to lead events of cascading failure of the power grid to chronic pain or epileptic seizure in the brain. Furthermore, networks provide a powerful model to understand and predict the properties of a diverse range of real-world complex systems. Recently, a multilayer network has been realized as a better suited framework for the representation of complex systems having multiple types of interactions among the same set of constituents. This article shows that by tuning the properties of one layer (network) of a multilayer network, one can regulate the dynamical behavior of another layer (network). By taking an example of a multiplex network comprising two different types of networked Kuramoto oscillators representing two different layers, this article attempts to provide a glimpse of opportunities and emerging phenomena multiplexing can induce which is otherwise not possible for a network in isolation. Here we consider explosive synchronization to demonstrate the potential of multilayer networks framework. To the end, we discuss several possible extensions of the model considered here by incorporating real-world properties.
\end{abstract}

\begin{keyword}
%% keywords here, in the form: keyword \sep keyword
modeling \sep multiplex network \sep explosive synchronization \sep Kuramoto oscillators
\end{keyword}

\end{frontmatter}

%% Start line numbering here if you want
%%\linenumbers

%%---------------------------------------------------------------------------------------------------------------------------------------------------------------------%%
%%---------------------------------------------------------------------------------------------------------------------------------------------------------------------%%
%% main text

%%--------------------------------------------------------------------------------------------------------------------------------------------------------------------%%
\section{Introduction}\label{Intro}
\noindent Network representation of a complex system, which includes nodes and connection between them called edges, provides insights about its structural formation and associated properties. Erd{\"o}s R{\'e}nyi's random network~\citep{Erdos1959}, Watts-Strogatz's small-world network~\citep{Watts1998} and Barabasi's scale-free network~\citep{Barabasi1999} are a few popular network topology in representing a variety of complex systems. In Erd{\"o}s R{\'e}nyi random network, each pair of $N$ nodes are randomly connected with a probability $p$. In small-world network, each node on a ring lattice having equal number of neighbors on either side, each edge is randomly rewired with a probability $p_r$. Scale-free network grows according to preferential attachment and its connectivity distribution follows power law distribution. The network framework has been tremendously successful model in understanding the capabilities and predicting the dynamical behaviour of diverse range of real-world complex systems. 
An underlying network structure permits the investigation of various stochastic and dynamical processes such as percolation~\citep{Kachhvah2012}, congestion~\citep{Litescu2015}, synchronization~\citep{Arenas2008}, chimera~\citep{Abrams2004}, disease epidemic~\citep{Youssef2011}, random walk and diffusion~\citep{Masuda2017} etc. on the corresponding complex systems.

In a number of real-world systems, the nodes can have multiple type of interactions among them
and hence correspondingly there may exist different connectivity patterns with the same set of 
neighbors. It has increasingly being realized that ignoring impact of multiplexing may result in wrong prediction of behavior of a given system. Such complex systems can be more precisely represented by a multilayer network architecture. A multiplex network consists of different layers containing the same set of nodes having one-to-one inter-layer links, with each layer having a distinct set of intra-layer links representing a distinct type of interactions among the nodes~\citep{Shakibian2016, Ding2018}. For instance, in a transportation network within a city, networks of connectivity of different modes of transport such as bus, metro and tram would form different layers and bus, metro and tram stops falling within a short radius would denote a node~\citep{Aleta2019}. A variety of studies have been carried out on top of multiplex network such as localization and optimization of synchronizability via single-layer rewiring~\citep{Pradhan2018,Pradhan2017}, multilayer protein-protein interaction networks of the cancerous tissues and various life stages in C. elegans~\citep{Aparna2017, Shinde2015}, and chimera and cluster synchronization etc. \citep{Jalan2016,Ghosh2016}. 
\begin{figure}[t!]
	\begin{center}
			\includegraphics[width=8cm, height=5cm]{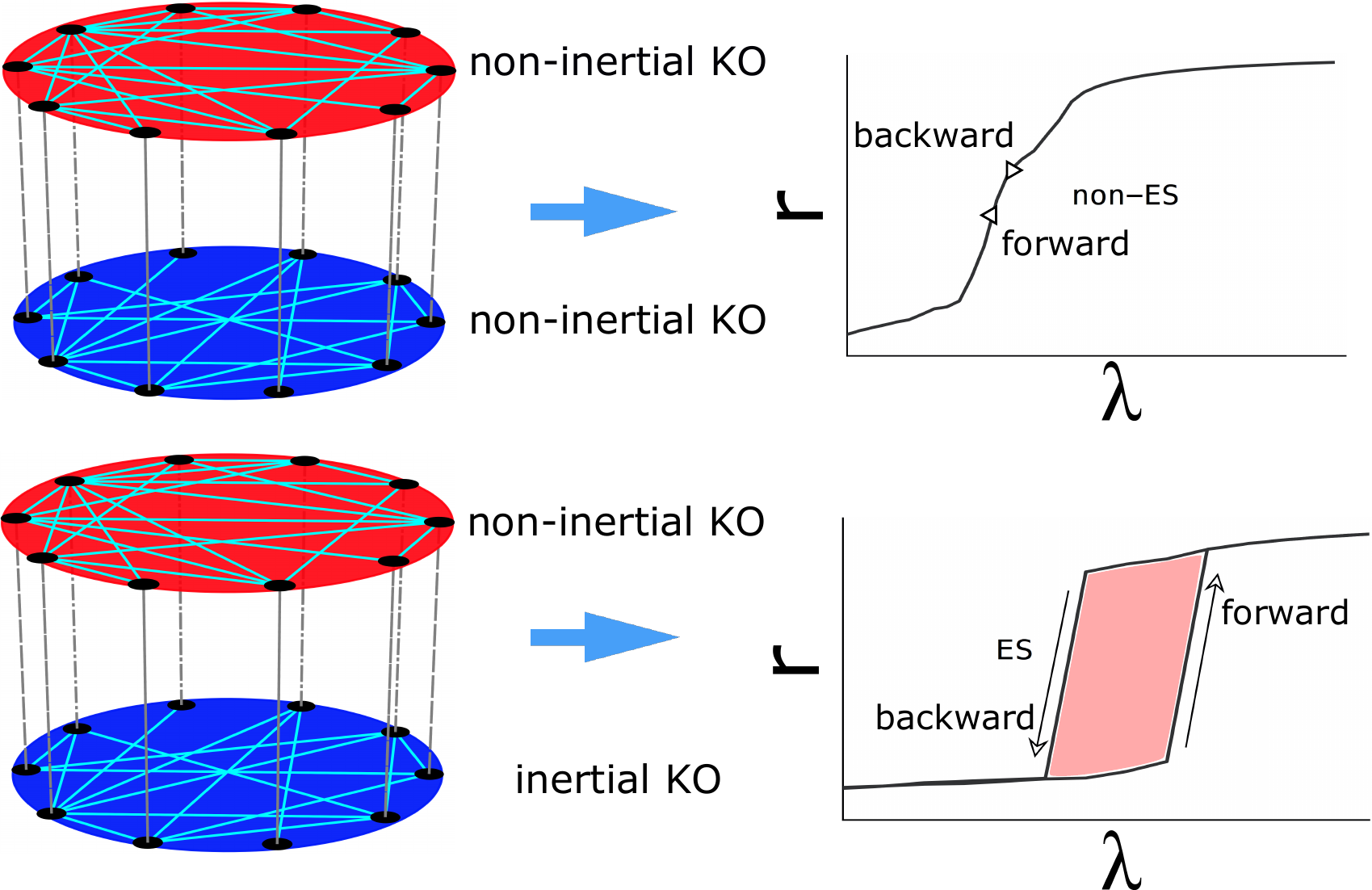}\\
	\end{center}
	\caption{Schematic representation of a multiplex network whose two layers follow the same dynamics, namely, non-inertial Kuramoto Oscillators (KO) and take up continuous route to achieve synchronous state. However, in other case when the dynamics follows non-inertial KO in one layer and inertial KO in other layer, both the layers then prompt to ES.}
	\label{fig:figure1}
\end{figure}

The advent of concept of complex network representing a variety of complex systems ranging from physical to biological systems, opened a new arena for network structure based study of synchronization. Synchronization is a process of unison of dynamics of initially perturbed interacting elements, for long considered to be a continuous (second-order) transition. However, recent studies have shown that under certain conditions, the phase transition to synchronized state in networked oscillators can be of first-order or discontinuous in nature. The discontinuous or explosive transition is a process in which all the nodes or a fraction of the nodes of a network precipitately join the largest synchronous cluster and then break off from it espousing an abrupt irreversible path, eventually forming a hysteresis loop. Taking into account special conditions such as a positive correlation of natural frequency with the oscillators' respective degree \citep{Gardenes2011} or coupling strength \citep{Zhang2013}, the presence of inertia \citep{Tanaka1997}, the presence of adaptive coupling \citep{Zhang2015, Danziger2019} or frequency mismatch \citep{Anil2019} have shown to lead to first-order or explosive transition to synchronous state in networked phase oscillators. Several real-world complex systems exhibit explosive synchronization, for example, cascade of blackouts in the power grids~\citep{Buldyrev2010}, congestion of the internet~\citep{Huberman1997}, chronic pain (Fibromyalgia) or epileptic seizures in the brain~\citep{Lee2018,Adhikari2013}, and hysteresis in the activation of embryonic Cdc2 cell~\citep{Pomerening2003}.

The study of occurrence of ES in complex system having multiple type of interactions becomes essential as to see how the interdependence of different process affect each other's behavior and can
lead to emerging phenomena not possible for a single network in isolation. Recent investigations 
have demonstrated that multilayer networks can exhibit ES for intra-layer adaptive coupling \citep{Zhang2015, Danziger2019}, intertwined coupling \citep{Nicosia2017}, in the presence of inertia \citep{Ajaydeep2017}, for delayed coupling while considering degree-frequency correlation \citep{Ajaydeep2019}, in the presence of inhibitory layer \cite{Vasundhara2019} and inter-layer frequency mismatch \citep{Anil2019}, etc. In the current article, to manifest the potential
of multiplex or multilayer network framework for modeling real-world complex systems, we consider
multiplex networks consisting of two layers with each layer having
having a different type of oscillators dynamics, namely Kuramoto oscillators~\citep{Kuramoto} and Kuramoto oscillators with inertia~\citep{Tanaka1997}. We show that the inertial Kuramoto layer can concurrently incorporate ES in the non-inertial Kuramoto layer which was incapable of manifesting ES in isolation.
We also discuss the impact of dynamical and structural parameters on the emerging ES in the multiplex network. Moreover, we put to test the robustness of the emerging ES against a variety of network topology of the multiplexed layers.

\begin{figure}[t!]
	\begin{center}
		\includegraphics[width=9cm, height=8cm]{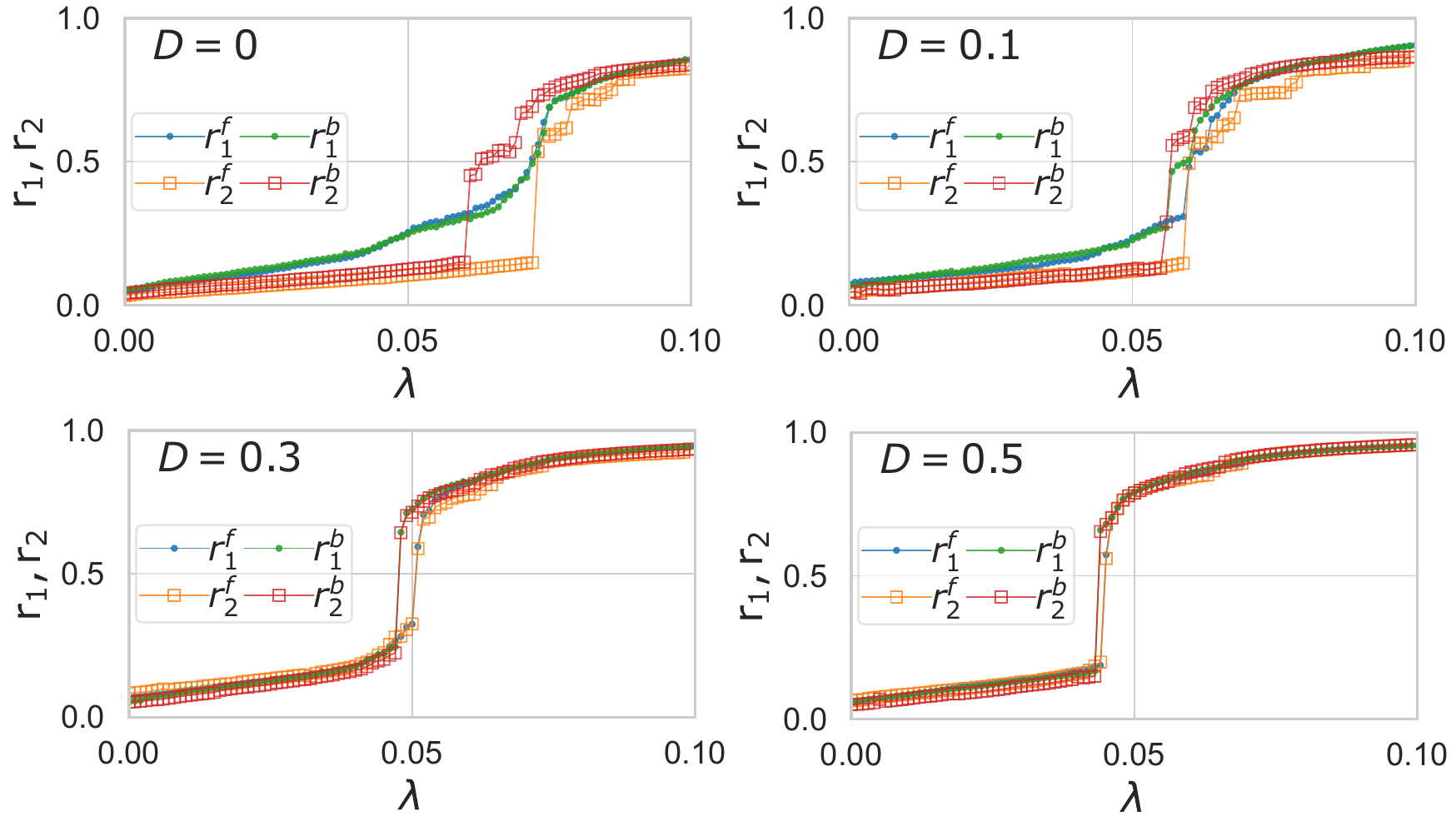}\\
		%\begin{tabular}{cc}
		%	\includegraphics[width=4.2cm, height=4cm]{km2km_1er2er_100n_m5_D0_op}& 
		%	\includegraphics[width=4.2cm, height=4cm]{km2km_1er2er_100n_m5_D01_op}\\
		%	\includegraphics[width=4.2cm, height=4cm]{km2km_1er2er_100n_m5_D03_op}& 
		%	\includegraphics[width=4.2cm, height=4cm]{km2km_1er2er_100n_m5_D05_op}\\
		%\end{tabular}
	\end{center}
	%\vspace{-1cm}
	\caption{Order parameter for the two layers as a function of coupling strength $\lambda$ for different values of $D$ when mass $m=5$.}
	\label{fig:figure2}
\end{figure}
%\begin{figure}[t!]
%	\begin{center}
%		\begin{tabular}{cc}
%			\hspace{-0.5cm}
%			\includegraphics[width=4.2cm, height=4cm]{km2km_1er2er_100n_m5_D0_op_iso}& 
%			\includegraphics[width=4.2cm, height=4cm]{km2km_1er2er_100n_m5_D1_op_mp}\\
%			%\includegraphics[width=4.2cm, height=4cm]{km2km_1er2er_100n_m5_D0_ef}& 
%			%\includegraphics[width=4.2cm, height=4cm]{km2km_1er2er_100n_m5_D1_ef}\\
%		\end{tabular}
%	\end{center}
%	\vspace{-1cm}
%	\caption{Order parameter of the non-inertial layer as a function of $\lambda$ in its (a) isolation and (b) multiplexing with inertial layer.}
%	\label{fig:figure2}
%\end{figure}

%%--------------------------------------------------------------------------------------------------------------------------------------------------------------------%%
\section{Model}\label{model}
\noindent Here we investigate the behavior of phase transition in interconnected networks governed by dissimilar dynamics namely, inertial and non-inertial Kuramoto oscillators as shown in Fig.\ref{fig:figure1}. Kuramoto oscillator is one of the most celebrated nonlinear dynamical model for investigating various complex physical and biological systems such as power-grid network~\citep{Dorflera2013} and neuronal network~\citep{Bansal2019}. We consider a multiplex network comprising two layers of $N$ nodes, with the nodes subject to first-order and second-order Kuramoto oscillator (KO) dynamics in the first and the second layer, respectively. The time-evolution of phases of interconnected nodes $\theta_l^i, l\in[1,2]$ in the multiplexed layers following non-inertial and inertial dynamics is respectively given by
\mathleft
\begin{subequations}
\begin{align} 
	\dot{\theta}_{1}^i = \omega^i_{1} +  {\lambda_1} \sum_{j=1}^N A^{ij}_{1} \sin(&\theta^j_{1}- \theta^i_{1}) \nonumber \\
			& + D \sin(\theta^i_{2}-\theta^i_{1}), \tag{1-a}\label{eq1}\\
	m\ddot{\theta}^i_2 + \dot{\theta}^i_{2} = \omega^i_{2} + {\lambda_2} \sum_{j=1}^N &A^{ij}_{2} \sin(\theta^j_{2}-\theta^i_{2}) \nonumber \\
	& + D \sin(\theta^i_{1}-\theta^i_{2}), \tag{1-b}\label{eq2}
\end{align}\label{model}
\end{subequations}
\mathcenter
where $\omega_l^i$ ($i=1,...,N$) denotes intrinsic frequency of $i^{th}$ node and mass $m$ denotes strength of inertia. Here $\lambda_1=\lambda_2=\lambda$ denotes intra-layer coupling strength among the nodes within the layer, and $D$ denotes the inter-layer coupling strength or multiplexing strength. The set of adjacency matrices ${\bf A}=\lbrace A_1, A_2\rbrace$ encodes connectivity information of the two multiplexed layers, where $A_l^{ij}$=1 if nodes $i$ and $j$ in layer $l$ are connected, and $A_l^{ij}$=0 otherwise.
\begin{figure}[t!]
	\begin{center}
		\includegraphics[width=9cm, height=11cm]{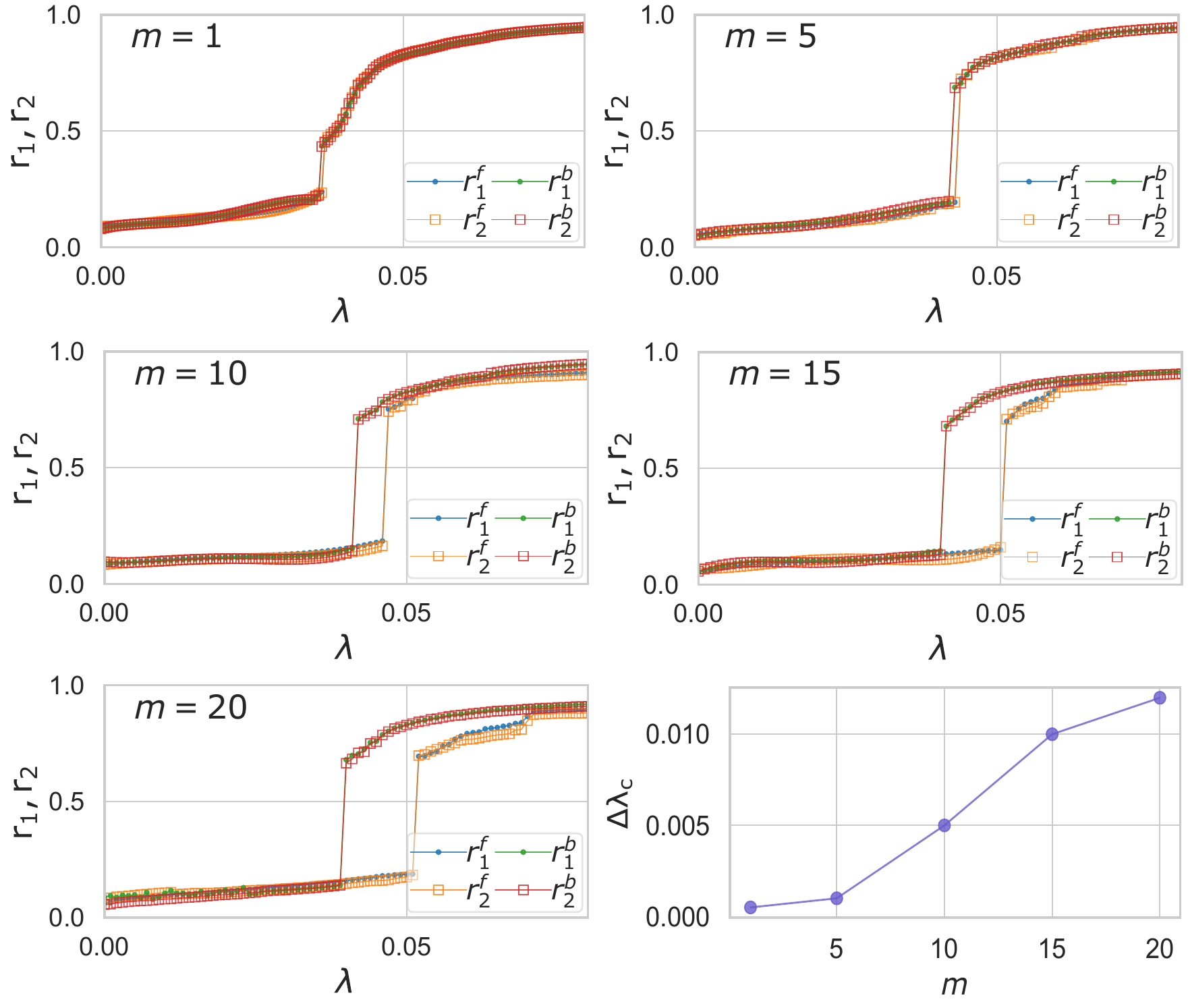}\\
		%\begin{tabular}{cc}
		%	\hspace{-0.6cm}
		%	\includegraphics[width=4.2cm, height=4cm]{km2km_1er2er_100n_m1_D1_op}&
		%	\includegraphics[width=4.2cm, height=4cm]{km2km_1er2er_100n_m5_D1_op}\\ 
		%	\includegraphics[width=4.2cm, height=4cm]{km2km_1er2er_100n_m10_D1_op}&
		%	\includegraphics[width=4.2cm, height=4cm]{km2km_1er2er_100n_m15_D1_op}\\ 
		%	\includegraphics[width=4.2cm, height=4cm]{km2km_1er2er_100n_m20_D1_op}&
		%	\includegraphics[width=4.2cm, height=3.85cm]{inertia_hyswidth}\\
		%\end{tabular}
	\end{center}
	%\vspace{-1cm}
	\caption{Order parameter for the two layers as a function of $\lambda$ for different values of $m$ when $D=1$. The hysteresis width $\Delta\lambda_c$ as a function of inertial strength $m$.}
	\label{fig:figure3}
\end{figure}

To capture the level of coherence in the multiplexed layers, we define order parameter $r_l$ for both the non-inertial and the inertial layer, with corresponding average phases $\psi_l$, as
\begin{equation}
    r_l(t)e^{\imath \psi_l} = \frac{1}{N}\sum_{j=1}^{N}e^{\imath\theta_l^j}.
\end{equation} \label{ordp_multi}
In the stationary state, $r_l=1$ corresponds to a completely synchronous state, i.e., insinuates the existence of the largest synchronous cluster while $r_l=0$ insinuates asynchrony in a layer $l$.  Also after eliminating an initial transients $t_r$ of the system state, if $T$ be the total time of temporal average, the effective frequency of a node $i$ in the layer $l$ is defined as 
\begin{equation}
\langle{\omega\rangle_l^i}=\frac{1}{T}\int_{t_r}^{t_r+T}\dot\theta_l^i(t)\mathrm{d}t.
\end{equation}\label{eff_freq}
The effective frequencies of all the nodes converge to a common average frequency in the synchronized state.

%Also the degree of coherence between each pair of connected nodes in a layer $l$ can be defined as follows~\cite{Vasundhara2019}
%\begin{equation}
%r^{ij}_l =A^{ij}_l\left |\lim_{T\rightarrow\infty}\frac{1}{T}\int_{t_r}^{t_r+T}e^{\imath\left[\theta^i_l(t)-\theta^j_l(t)\right]}\mathrm{d}t\right|\;.
%\label{eq:rij}
%\end{equation}
%$r^{ij}_l=1$ if connected nodes $(i,j)$ are in complete synchrony and $r^{ij}_l=0$ if they are incoherent. If  $2N_c$ be the total count of existing connections in a network layer, the degree of local clustering among the nodes in a network can then be captured by
%\begin{equation}
%r^{link}_l=\frac{1}{2N_c}\sum_{i}\sum_{j} r^{ij}_l.
%\label{r_link}
%\end{equation}

%%--------------------------------------------------------------------------------------------------------------------------------------------------------------------%%
\section{Results}\label{results}
\noindent For our numerical investigation, we first consider a multiplex network of two Erd{\"o}s R{\'e}nyi (ER) random networks~\citep{Erdos1959}, each having $N=100$ nodes and average connectivity $\langle k\rangle_1=\langle k\rangle_2=10$, otherwise mentioned elsewhere. The different realizations of intrinsic frequencies  (initial phases) of the nodes in two layers are uniformly randomly sampled from the interval $\omega^i_l\in[-0.5,0.5)$ ($\theta^i_l\in[0,2\pi)$).

Dynamics of the multiplex system (Eq.~(\ref{model})) is evolved for sufficiently long time
($5 \times 10^4$ time steps) by employing RK4 method with time-step of $\mathrm{d}t = 0.01$ so as to arrive at stationary state $r_l(t)\simeq r_l$ after discarding initial transients. To explore the existence of hysteresis, we compute order parameter for both the layers in forward continuation ($r^f$) and backward continuation ($r^b$) of $\lambda$. In the forward continuation, $\lambda$ is increased adiabatically, i.e., while increasing $\lambda$ from $0$ onwards to some $\lambda$ (corresponding to synchronized state) in the steps of $\mathrm{d}\lambda$, the steady phases obtained for $\lambda+n\mathrm{d}\lambda$ is fed as initial state for $\lambda+(n+1)\mathrm{d}\lambda$ and so on. In the similar fashion in the backward continuation, $\lambda$ is decreased adiabatically, i.e., the steady phases for $\lambda+n\mathrm{d}\lambda$ is fed as initial state for $\lambda+(n-1)\mathrm{d}\lambda$ while decreasing $\lambda$ from its value corresponding to synchronized state to $0$.

%The behavior of order parameter of a non-inertial (KO) layer, as a function of coupling strength $\lambda$, in its isolation and its multiplexing ($D=1$) with inertial (KO) layer is depicted in Fig.~\ref{fig:figure2}. It unveils that the non-inertial layer which takes up continuous route ($r$) to achieve synchronous state in its isolation, upon multiplexing with inertial layer adopts explosive (discontinuous) ($r_1$) route to synchronization. Hence, multiplexing of a dynamical network with some other dynamical network supporting ES, triggers ES in the former network layer. Next, we investigate the impact of dynamical and structural features of the employed multiplex model on the emerging ES transition.

\subsection{Emergence of ES through multiplexing}
This section demonstrates that how the non-inertial KO layer taking up continuous route to achieve synchronous state in its isolation, upon multiplexing with inertial KO layer espouses explosive (discontinuous) route to synchronization. To comprehend this, behavior of the order parameter
for the two layers in the forward ($r^f$) and the backward continuation ($r^b$) of $\lambda$ for different multiplexing strength $D$ is illustrated in Fig.~\ref{fig:figure2}. In the absence of 
inter-layer coupling, i.e., $D=0$, the non-inertial (first-order) and the inertial (second-order) KO dynamical layer espouses the continuous ($r_1$) and the discontinuous ($r_2$) route to synchronized state, respectively. The inertial layer forms a hysteresis loop for the forward ($r^f$) and the backward ($r^b$) continuation of transition. However in the presence of a weak multiplexing ($D=0.1$), a small fraction of the non-inertial nodes concurrently becomes part of the largest synchronized cluster due to the impression of frustration from the inertial nodes (layer). Thus, the non-inertial layer starts sporting a short abrupt (ES) transition with an associated hysteresis. For stronger $D$, the non-inertial layer concurrently follows the ES transition of the inertial layer all along. As the multiplexing strength $D$ gets stronger, the impression of suppression on non-inertial nodes from inertial nodes gets reinforced. Consequently, a larger fraction of the non-inertial nodes gets synchronized simultaneously at a rather lower forward coupling strength giving rise to a ES transition with an increased jump size. On this account, increase in $D$ leads to increase in discontinuous jump size and fall in forward critical coupling strength.

For a fixed $m$, while decreasing $\lambda$ starting from a synchronous state ($r^b\simeq1$), the impact of suppression on the formation of giant cluster due to a weaker $D$ is weak, thus even at a higher value of $\lambda$ (early in the backward continuation) the existing frustration among the nodes would die out and the nodes would break off from the synchronous cluster rather gradually. Nevertheless, a stronger $D$ would yield strong frustration among the nodes, hence the nodes would tend to remain in the synchronous cluster until a rather lower critical value of $\lambda$ (later in the backward continuation) is reached. Thus for a fixed $m$, hysteresis width is decreased with an increase in the multiplexing strength $D$ \citep{Ajaydeep2017}.

\subsection{Controlling ES through inertial strength $m$}
Next, we show that how the inertial strength $m$ controls the behavior of emerging ES transition.
Fig.~\ref{fig:figure3} depicts the behavior of order parameter for the multiplexed layers as a function of  $\lambda$ for different values of the inertial strength. For very weak $m$, the two layers espouse second-order transition. Inertial strength $m$ induces frustration among the nodes and suppresses synchronization until a minimum value of $\lambda$ is reached, which overcoming the suppression pulls all the nodes from asynchronous state to synchronous state simultaneously giving rise to an abrupt jump in the order parameter. For $m=1$, the discontinuous jump size is very small as weaker frustration compels mere a few nodes to synchronize simultaneously. With a further increase in $m$ leads to an increase in frustration, accordingly a larger fraction of the nodes abruptly synchronizes at even higher critical $\lambda$. On this account, value of the forward critical coupling strength increases as $m$ increases.

In the backward continuation, system initially starts from $r^b\simeq1$ state where $\lambda$ dominates over the frustration in both the layers so that all the nodes remain attached to the largest synchronous cluster. As $\lambda$ keeps decreasing, its dominance over frustration starts subsiding steadily and accordingly a very few nodes starts breaking off from the largest cluster. This situation persists until a critical value of $\lambda$ is met at which all of sudden frustration completely dominates over the intra-layer coupling strength. This leads to nearly complete desynchronization of the nodes in both the layers, which is corroborated by steep fall in the order parameter. For that reason, an increase in the inertial strength $m$ yields an increase in the frustration among the nodes, thereby lowering the value of the backward critical $\lambda$ at which abrupt desynchronization takes place.  On this account, the hysteresis width $\Delta\lambda_c$ which is equal to the difference between the backward and the forward critical coupling strength, gradually increases as inertial strength $m$ increases (see $\Delta\lambda_c-m$ profile in Fig.~\ref{fig:figure3}). Hence, the order parameter for the two layers concurrently espouses explosive route to synchronization with associated hysteresis for a significant range of the inertial strength.
\begin{figure}[t!]
	\begin{center}
		\includegraphics[width=9cm, height=8cm]{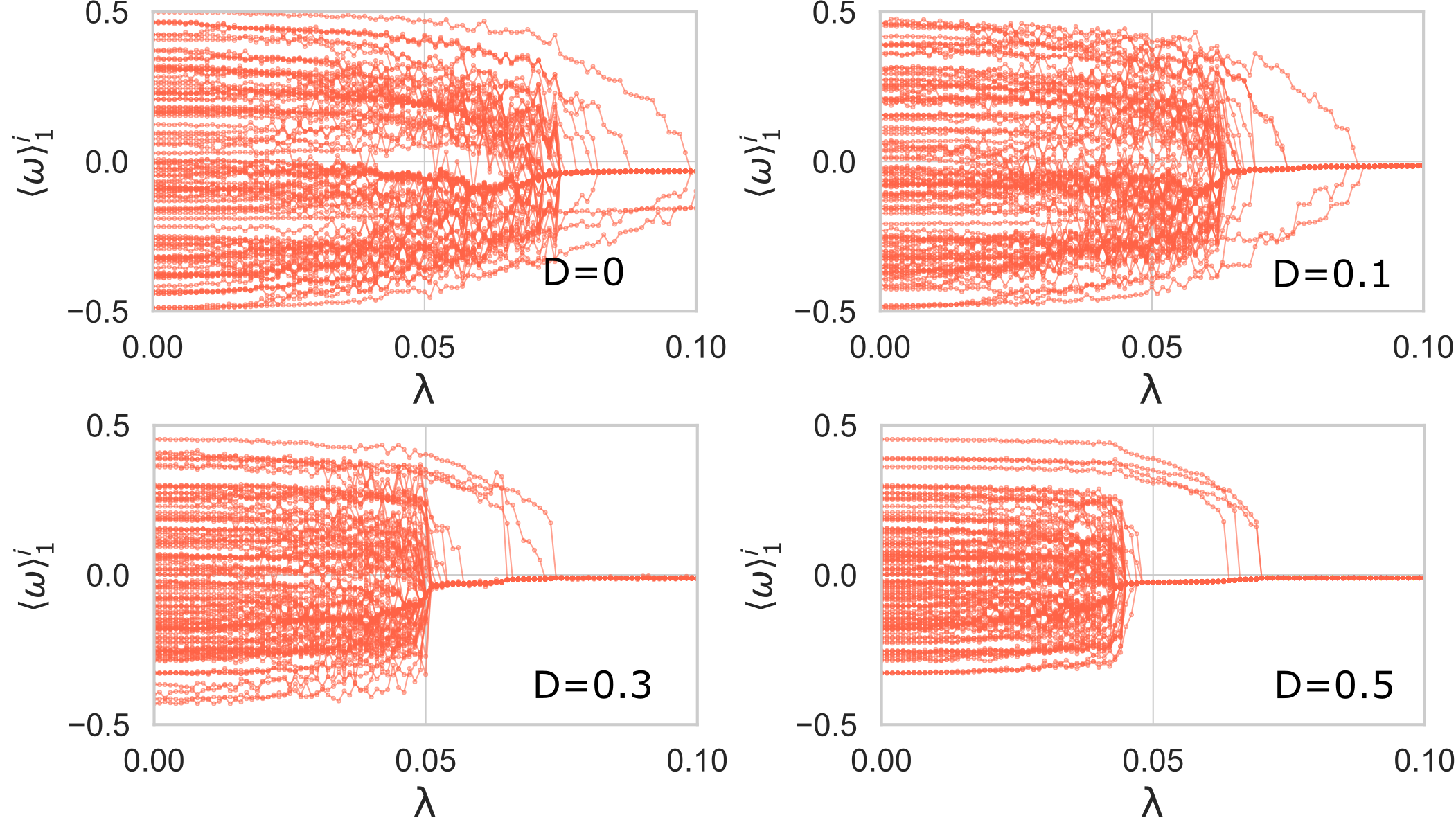}\\
		%\begin{tabular}{cc}
			%\hspace{-1cm}
			%\includegraphics[width=4.2cm, height=4cm]{km2km_1er2er_100n_m5_D01_rl}& 
			%\includegraphics[width=4.2cm, height=4cm]{km2km_1er2er_100n_m5_D05_rl}\\
			%\hspace{-1cm}
			%\includegraphics[width=4.2cm, height=4cm]{km2km_1er2er_100n_m5_D01_ef}& 
			%\includegraphics[width=4.2cm, height=4cm]{km2km_1er2er_100n_m5_D05_ef}\\
		%\end{tabular}
	\end{center}
	%\vspace{-1cm}
	\caption{$\langle\omega\rangle_1^i$ of non-inertial layer of the multiplex network as a function of $\lambda$ corresponding to $m=5$ and different values of $D$.}
	\label{fig:figure4}
\end{figure}

\subsection{Microscopic dynamics behind origin of ES}
To understand the underlying microscopic dynamics taking place behind the origin of ES, we study the behavior of effective frequencies for the non-inertial layer in Fig.\ref{fig:figure4}. For $D=0$, $\langle\omega\rangle^i_1$ of the non-inertial layer  gradually converge to a common mean frequency leading to continuous transition. Further, even a very weak $D=0.1$ is capable of inducing ES with short jump (see Fig.\ref{fig:figure2}), which can be corroborated by the behavior of $\langle\omega\rangle^i_1$ suggesting that only a small fraction of the nodes abruptly gets synchronized at critical coupling strength. As inter-layer coupling strength gets more significant ($D=0.3\ \mbox{and then}\ 0.5$), a larger fraction of the nodes gets synchronized at rather smaller critical $\lambda$ by reducing the remaining fraction of the asynchronous nodes which synchronize gradually later beyond the 
critical $\lambda$. This leads to ES with profound discontinuous jump. In this fashion, collective behavior of effective frequencies of a fraction the nodes accounts for the increase in the jump size with increase in $D$. 

\subsection{Robustness of ES against network topology}
\begin{figure}[t!]
	\begin{center}
		\includegraphics[width=9cm, height=9cm]{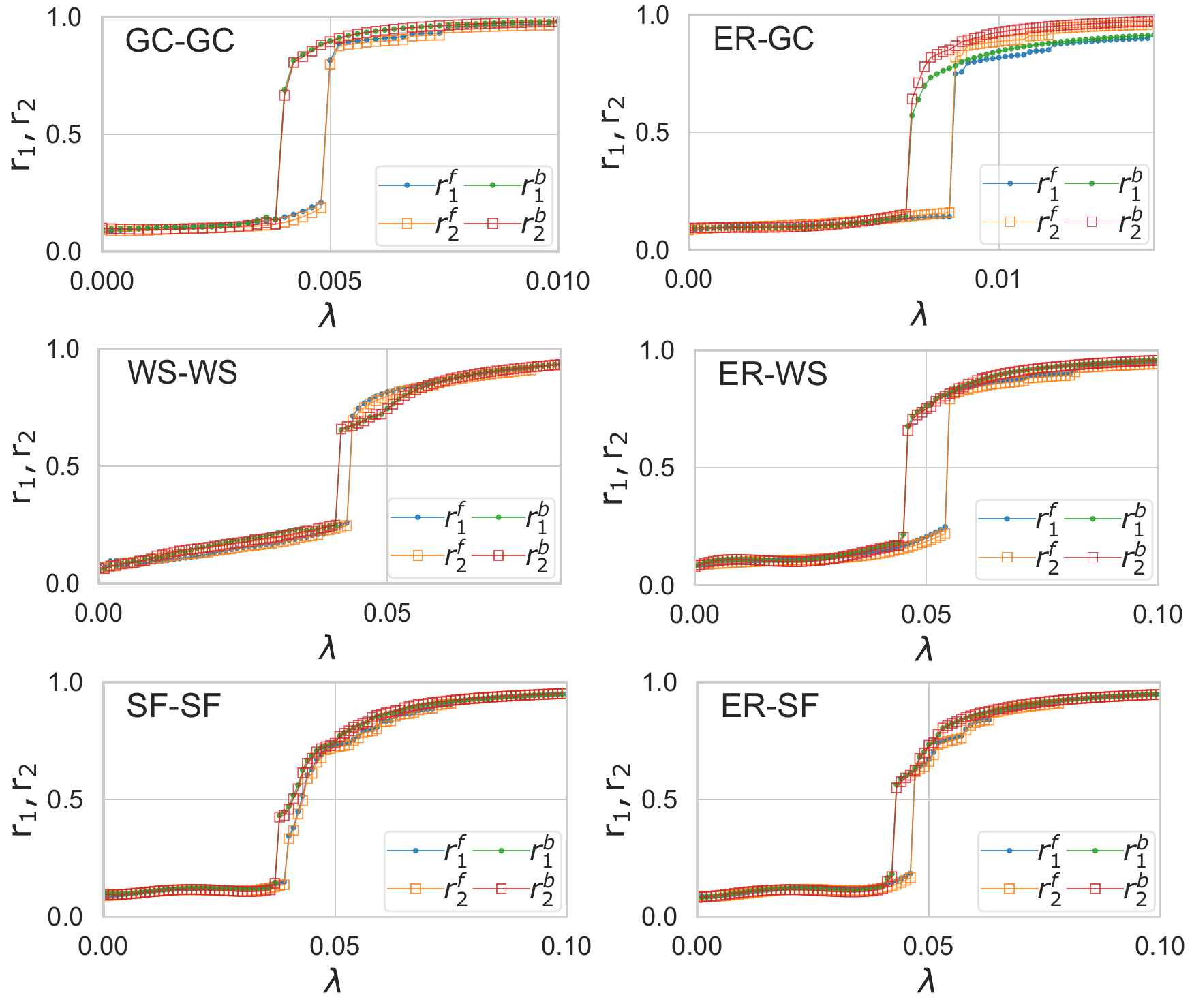}\\
	\end{center}
	%\vspace{-1cm}
	\caption{$r^f$ and $r^b$ as a function of $\lambda$ for multiplex networks of different topology with $m=10$ and $D=1$. Panels in first column correspond to multiplex network comprising two GC, WS and SF networks. Panels in second column correspond to multiplex network comprising first layer of ER topology and second layer of GC, WS and SF topology. Note that $N=100$ nodes in each layer $l$ sporting either ER or WS or SF topology have average degree $\langle k\rangle_l=10$. Graphs are generated with connection probability $0.1$ in ER, rewiring probability $0.05$ in WS and addition of each new node with $5$ edges in SF.}
	\label{fig:figure5}
\end{figure}

Here we put to the test the robustness of emergent ES transition against a variety of topology selected for the two layers by exploring the behavior of order parameter as a function of $\lambda$ (see Fig.~\ref{fig:figure5}). Panels in the first column of Fig.~\ref{fig:figure5} correspond to the multiplex networks consisting of two layer with each layer represented by globally connected (GC), small-world (SW) and scale-free (SF) topology. The employed multiplex framework with contrasting dynamics supports ES transition even for different types of the network topology.  
Further, we check the robustness of employed model for dissimilar network topology for the
different layers (second column, Fig.~\ref{fig:figure5}). By fixing one layer to the ER topology while the other layer being tested against the GC, WS and SF topology. This dissimilar topology test also endorses ES transition route for the employed model. Hence, the proposed model is robust enough in inducing ES transition with profound hysteresis in the multiplexed layers with a variety of similar or dissimilar network topology.

\section{Analytical treatment}
In order to obtain analytical expression for order parameter, for simplicity, we take into account a multiplex network comprising two GC layers $l\in[1,2]$ so that $A_l^{ij}=1/N$ in model Eq.(\ref{model}). Also we consider a symmetric unimodal frequency distribution $g(\omega_l)=g(-\omega_l)$ for both the layers. Further, we consider the continuum limit ($N \rightarrow \infty$) of the model Eq.(\ref{model}) by defining a density function $\rho_l(\omega_l,\theta_l,t)$, which is in fact the fraction of oscillators having frequency $\omega_l$ with values of phases lying between $\theta_l$ and $\theta_l+d\theta_l$ at instant $t$. $\rho_l(\omega_l,\theta_l,t)$ is normalized as $\int_0^{2\pi}  \rho_l(\omega_l,\theta_l,t) d\theta_l=1$ for all $\omega_l$ and all $t$, and its evolution is governed by the continuity equation $\partial\rho_l/\partial t+ \partial(\rho_l v_l)/\partial\theta_l=0$, where $v_l$ is the velocity of oscillators given by $\dot\theta_l$ from Eq.(\ref{model}). 

\begin{figure}[t!]
	\begin{center}
		\includegraphics[width=8cm, height=5.5cm]{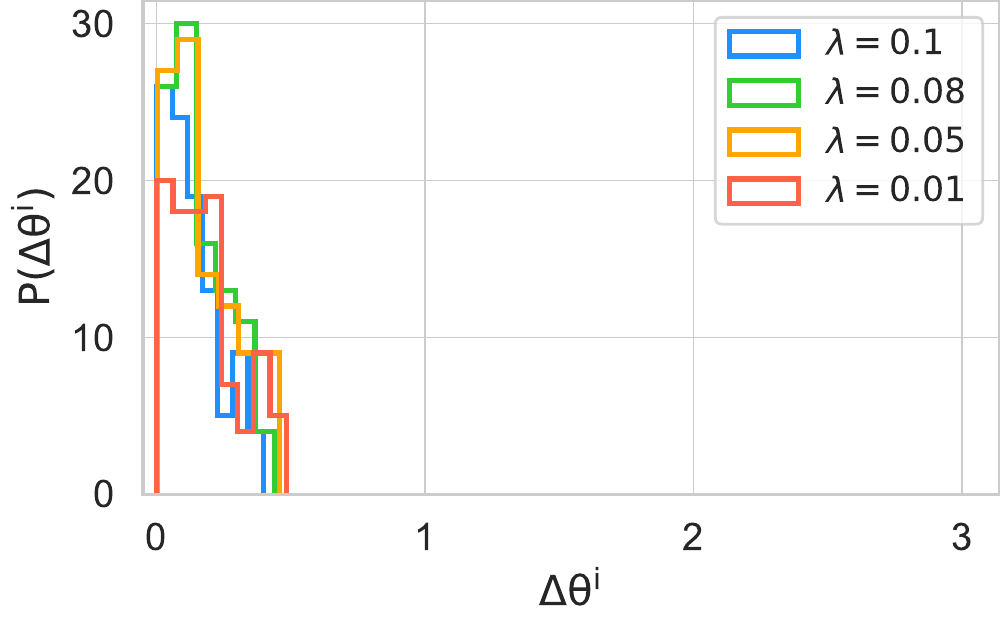}\\
	\end{center}
	\vspace{-0.5cm}
	\caption{Distribution of $\Delta\theta^i=|\theta_2^i-\theta_1^i|$ corresponding to different $\lambda$ for multiplex network comprising ER-ER layers having uniform distribution for natural frequencies, $m=10$ and $D=1$.}
	\label{fig:figure6}
\end{figure}

The order parameter given by Eq.(\ref{ordp_multi}) can be rewritten in the continuum limit as 
\begin{equation}\label{op_cont}
r_le^{i\psi_l} = \int d\omega_l \int d\theta_l g(\omega_l) \rho_l(\omega_l,\theta_l,t) e^{\imath\theta_l}.
\end{equation}
Eq.(\ref{model}) can be written in mean-field form using Eq.(\ref{ordp_multi}) or Eq.(\ref{op_cont}) in the continuum limit as
\mathleft
\begin{align} \label{model_mf}
	\dot{\theta}_{1}^i = \omega^i_{1} +  {\lambda}r_1 \sin(\psi_{1}- \theta^i_{1}) 
			+ D \sin(\theta^i_{2}-\theta^i_{1}),\nonumber \\
	m\ddot{\theta}^i_2 + \dot{\theta}^i_{2} = \omega^i_{2} + {\lambda}r_2 \sin(\psi_{2}-\theta^i_{2}) - D \sin(\theta^i_{2}-\theta^i_{1}).
\end{align}
\mathcenter
From numerical simulations it turns out that the distribution of $\Delta\theta^i=|\theta_2^i-\theta_1^i|$, the difference between phases of the interconnected nodes in layers $1$ and $2$, for any $\lambda$ follows Normal (unimodal) distribution with its mean centered at $0$ and standard deviation $\Delta\theta(\lambda)$ as shown in Fig.\ref{fig:figure6}. Since $\Delta\theta(\lambda)$ is considerably small, hence coupling term consisting of inter-layer coupling strength and interaction between the mirror nodes can be considered a constant for any $\lambda$, i.e., $D \sin(\theta^i_{2}-\theta^i_{1})=D\sin(\Delta\theta)\simeq D.\Delta\theta=\sigma$ (say). Also without a loss of generality, one can always set $\psi_1=0$ and $\psi_2=0$. Hence, the model Eqs.(\ref{model_mf}) can be rewritten in simplified mean-field form as 
\begin{align} \label{model_new}
	\dot{\theta}_{1}^i = \Omega^i_{1} -  {\lambda}r_1 \sin\theta^i_{1},\nonumber \\
	m\ddot{\theta}^i_2 + \dot{\theta}^i_{2} = \Omega^i_{2} - {\lambda}r_2 \sin\theta^i_{2},
\end{align}
where $\Omega_1^i=\omega_1^i+\sigma$, and $\Omega_2^i=\omega_2^i-\sigma$. In this fashion, $\sigma(\lambda)=D.\Delta\theta(\lambda)$ accounts for the maximum possible inter-layer contribution in the evolution of phases in either layer.

Since we are interested in steady state solutions, i.e., $\partial\rho_l/\partial t=0$, hence time-independent solution of $\rho_l(\Omega_l,\theta_l,t)$ is given by $\partial(\rho_l v_l)/\partial\theta_l=0$ as following
\begin{equation}\label{density_fn}
 \rho_l(\Omega_l,\theta_l)=\left\{
                        \begin{array}{ll}
                         \delta(\theta_l - \arcsin(\frac{\Omega_l}{\lambda r_l})) & \mbox{ if } \frac{|\Omega_l|}{\lambda r_l} \le 1 \\
                         \frac{A(\Omega_l)}{|\Omega_l-\lambda r_l\sin\theta_l|} &  \mbox{ if } \frac{|\Omega_l|}{\lambda r_l} > 1,
                        \end{array}
                       \right.
\end{equation}
where is $A(\Omega_l)$ the normalization factor. Now expression for $r_l$ can be obtained by inserting Eq.(\ref{density_fn}) into Eq.(\ref{op_cont}) and dividing the integral over $\Omega_l$
\begin{align}\label{op_para}
r_l = \left[\int_{-\Omega_l^p}^{\Omega_l^p} d\Omega_l + \int_{-\infty}^{-\Omega_l^p} d\Omega_l + \int^{\infty}_{\Omega_l^p} d\Omega_l\right] \nonumber \\ 
\int_0^{2\pi} d\theta_l g(\Omega_l\mp\sigma) \rho_l(\Omega_l,\theta_l) e^{\imath\theta_l},
\end{align}
where $g(\Omega_l-\sigma)$ and $g(\Omega_l+\sigma)$ correspond to layers $l=1$ and $l=2$, respectively. Also $\Omega_l^p$ is the threshold limit for either layer $l$.

The contribution from locked oscillators obeying $-\Omega_l^p\le\Omega_l\le\Omega_l^p$ can be determined by inserting $\rho_l(\Omega_l,\theta_l)$ from Eq.(\ref{density_fn}) into Eq.(\ref{op_para})
\begin{equation}
 r_l^{L} = \int_{-\Omega_l^p}^{\Omega_l^p}d\Omega_l\ g(\Omega_l\mp\sigma)\exp\left[i \arcsin\left(\frac{\Omega_1}{\lambda r_l}\right)\right].
\end{equation} 
After some mathematical working out, the real contribution to order parameter from locked oscillators for either layer can be expressed as
\begin{equation}\label{op_lock}
 r_l^{L} = \int_{-\Omega_l^p}^{\Omega_l^p}d\Omega_l\ g(\Omega_l\mp\sigma) \sqrt{1-\left(\frac{\Omega_l}{\lambda r_l}\right)^2}.
\end{equation} 

\subsection{Order parameter $r_1$ for non-inertial layer}
For forward continuation of the non-inertial layer $1$, Eq.(\ref{density_fn}) yields integral limit $\Omega_1^p{=}\lambda r_1$. If we consider $g(\omega_l)$ to follow Lorentzian or Gaussian distribution for which $g(\omega_l){=}g(-\omega_l)$ and $\rho_l(-\omega_l,\theta_l{+}\pi)=\rho_l(\omega_l,\theta_l)$, the contribution from drifting oscillators to layer $1$ obeying $\Omega_1{<-} \lambda r_1$ and $\Omega_1 {>} \lambda r_1$ vanishes as 
\mathleft
\begin{align}
 r_1^{D}=\left[\int_{-\infty}^{-\lambda r_1} d \Omega_1{+}\int_{\lambda r_1}^{\infty}d \Omega_1 \right]
 g(\Omega_1{-}\sigma)\rho_1(\Omega_1,\theta_1)e^{i\theta_1}\nonumber \\
{=}\int_{\lambda r_1}^{\infty}g(\Omega_1{-}\sigma)\left[\frac{A(\Omega_1)e^{i\theta_1}}{\Omega_1{-}\lambda r_1\sin\theta_1}+\frac{A(\Omega_1)e^{i\theta_1}}{\Omega_1{+}\lambda r_1\sin\theta_1}\right]{=}0.
\end{align}\label{for-0}
\mathcenter
Hence for the non-inertial layer, only the locked oscillators would contribute to $r_1$, i.e., $r_1 = r_1^{L} + r_1^{D}=r_1^L$. From Eq.~(\ref{op_lock}), $r_1$ for Lorentzian distribution $g(\omega_1)=\frac{d}{\pi(\omega_1^2+d^2)}$ can be mathematically simplified as
\begin{align}\label{op1}
r_1 = \frac{\lambda r_1 d}{\pi} \int_{-1}^1 dx \frac{\sqrt{1-x^2}} {(\lambda r_1 x-\sigma)^2+d^2}.
\end{align}

\subsection{Order parameter $r_2$ for inertial layer}
\begin{figure}[t!]
	\begin{center}
		\includegraphics[width=8cm, height=6cm]{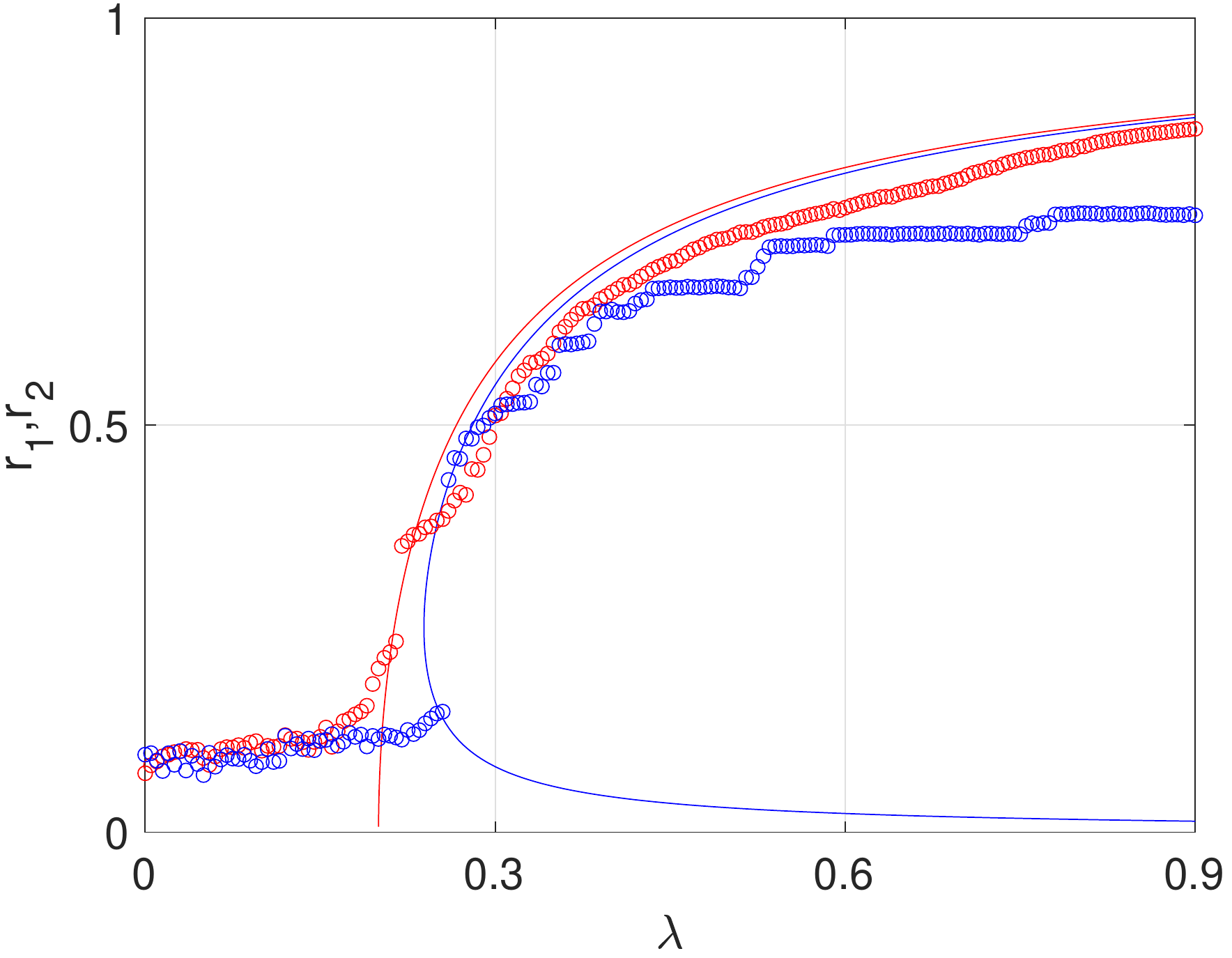}\\
	\end{center}
	\vspace{-0.5cm}
	\caption{Order parameters $r_1$ (red color) and $r_2$ (blue color) for two multiplexed GC layers. The circles denote simulation data while solid lines denote analytical predictions computed from Eq.(\ref{op1}) (red one) and Eq.(\ref{op2}) (blue one). The simulations are performed for $N=100$ nodes in each layer, inertia $m=15$, and location$=0$ and scaling $d=0.1$ of Lorentzian distribution.}
	\label{fig:figure7}
\end{figure}

To derive an expression for order parameter for inertial layer $2$, we carry out some mathematical manipulations inspired from Tanaka et.al \citep{Tanaka1997} using Eq.(\ref{density_fn}) and Eq.(\ref{op_para}) for the contribution of drifting oscillators. The contribution from drifting oscillators to $r_2$ is expressed as~\citep{Tanaka1997}
\begin{align}
r_2^{D} = -m\lambda r_2 \int_{\Omega_2^p}^{\infty}\frac{g(\Omega_2+\sigma)}{(m\Omega_2)^3}\mathrm{d}\Omega_2,
\end{align}
where $\Omega_2^p=\frac{4}{\pi}\sqrt{\lambda r_2/m}$ is integral limit for forward contribution of the inertial layer. If $g(\omega_2)$ follow Lorentzian distribution of form $g(\omega_2)=\frac{d}{\pi(\omega_2^2+d^2)}$, the contribution from drifting oscillators can then be expressed as
\begin{equation}
r_2^{D} = -\frac{\lambda r_2d}{m^2\pi} \int_{\Omega_2^p}^{\infty}\frac{1}{\Omega_2^3[(\Omega_2+\sigma)^2+d^2]}\mathrm{d}\Omega_2,
\end{equation}
Now from Eq.(\ref{op_lock}), the contribution from locked oscillators to $r_2$ can be expressed as
%\begin{equation}
%r_2^{L} = \int_{-\Omega_2^p}^{\Omega_2^p}d\Omega_2\ g(\Omega_2+\sigma) \sqrt{1-\left(\frac{\Omega_2}{\lambda r_2}\right)^2}.
%\end{equation} 
\begin{align}\label{op_lock2}
r_2^L = \frac{\lambda r_2 d}{\pi} \int_{-x^p}^{x^p} dx \frac{\sqrt{1-x^2}} {(\lambda r_2 x+\sigma)^2+d^2},
\end{align}
where $x^p=\frac{4}{\pi}\sqrt{1/(m\lambda r_2)}$.
Hence, order parameter for the inertial layer $2$ is determined by 
\begin{equation}\label{op2}
r_2 = r_2^{L}+r_2^{D}.
\end{equation}
Next, we compare the obtained analytical predictions for $r_1$ and $r_2$ respectively given in Eq.(\ref{op1}) and Eq.(\ref{op2}) with their respective numerical estimations as shown in Fig.\ref{fig:figure7}. It is obvious that analytical predictions made match quite well with respective numerical estimations, however analytic curve for $r_2$ seem to deviate a bit at larger values of $\lambda$.

%%--------------------------------------------------------------------------------------------------------------------------------------------------------------------%%
\section{Conclusion}\label{Conc} 
\noindent ES in multiplex networks is known to arise from intra-layer adaptive coupling, intra-layer inhibitory coupling, inertia, degree-frequency correlation etc.  All these works have considered
one type of local dynamics for individual nodes in both the layers. Here, we have demonstrated that ES can be originated in a multiplex network with layers having distinct local dynamics and while one layer sporting ES transition in its isolation. We have demonstrated robustness of the
the proposed multiplex network model in yielding ES by its layers for a variety of network topologies. The proposed model in the presence of a rather strong inertia supports ES in the multiplexed layers even for very low value of the multiplexing strength. The strength of inertia avails the required suppression in the formation of the largest synchronous cluster. Hence the strength of inertia, i.e., $m$ determines the onset of ES and width of the hysteresis.

\section{Future Perspectives}
\noindent This study would advance our understanding apropos the dynamical behavior of interconnected elements of physical or biological systems, which are governed by different type of dynamics corresponding to different type of interactions. This section provides a few future directions and open problems in the framework of the multilayer model proposed here.
 
This article has restricted to the phenomena of ES, however the same multiplex framework can be 
used to understand and model various other dynamical features such as cluster synchronization, relay synchronization, chimera, percolation, epidemic spreading, etc. displayed by complex systems.
Further, the multiplex framework may provide an alternative method for controlling a system. In traditional control theory, a desired state of a dynamical system can be achieved by providing suitable inputs to a few nodes \cite{Liu2011}, whereas in multiplex framework one can control a networked system or dynamical properties of one layer by appropriate multiplexing with another layer. In fact, it has been demonstrated that such control by appropriate multiplexing may not be only possible for certain systems but also can provide a more cost effective control as compared to traditional optimization methods~\cite{Dwivedi2017}. 
\begin{figure}[t!]
	\begin{center}
		\includegraphics[width=8cm, height=4cm]{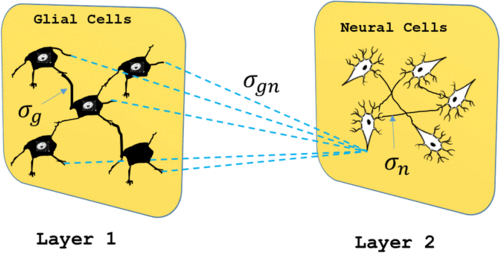}\\
	\end{center}
	%\vspace{-0.5cm}
	\caption{Schematic diagram of a bidirectional multilayer network illustrating intra-layer $\sigma_g$ and $\sigma_n$ coupling links among glial cells and neuronal cells, respectively, along with inter-layer coupling $\sigma_{gn}$. Reprinted with permission from S. Makovkin \emph{et al.}  Phys. Rev. E 96, 052214 (2017). Copyright 2017 American Physical Society.}
	\label{fig:figure8}
\end{figure}

So far, many important dynamical phenomena such as synchronization, chimera etc. have been 
investigated on multiplex networks with each layer modeled by the same coupled dynamics. 
However, there exists several real-world complex systems having multilayer architecture, where one layer is approximately modeled by one type of coupled dynamics and another layer by other type of coupled dynamics.
This article provides a framework to extend the studies carried on the similar dynamics on each layer of multiplex networks by taking different dynamical systems representing different layers. 
In the following we enlist a few extension of the model system taken here by incorporating properties of real-world complex systems.

\subsection{Impact of Phase lag} 
The article only concentrates on phase lag $\alpha$ being zero, whereas by introducing a phase lag in one or both the layers can lead to rich dynamical patterns. The inclusion of phase lag in both the multiplexed layers can be accommodated as following 
\mathleft
\begin{align}
	\dot{\theta}_{1}^i = \omega^i_{1} +  {\lambda} \sum_{j=1}^N A^{ij}_{1} \sin(\theta^j_{1}- &\theta^i_{1}-\alpha)\nonumber \\
	 &+ D \sin(\theta^i_{2}-\theta^i_{1}),\nonumber \\
\begin{split}
	m\ddot{\theta}^i_2 + \dot{\theta}^i_{2} = \omega^i_{2} + {\lambda} \sum_{j=1}^N A^{ij}_{2} &\sin(\theta^j_{2}-\theta^i_{2}-\alpha)\nonumber\\
 &+ D \sin(\theta^i_{1}-\theta^i_{2}).\nonumber
 \end{split}
\end{align}\label{model2} 
\mathcenter
The presence of non-zero phase lag(s) has been shown to be crucial for obtaining chimera patterns in single layer as well as in multiplexed layers \cite{Frolov2018}, and to control the nature of emergent ES transition~\cite{Khanra2018}. However all these investigations are restricted to both the layer having the same local dynamics and impact of inclusion of phase in multiplex networks having dissimilar dynamics on coupled dynamical evolution, particularly on ES,
remains an open problem.

\subsection{Inclusion of communication delays}
Communication delays are naturally present in various real-world complex systems due to the finite propagation speed of signal transmission, and have been demonstrated to play an important role in deciding the combined dynamical evolution of such coupled systems. For example, delays are an integral part for the processing of brain signals to provide a combined and coherent perception of the outer world. The inclusion of time-delay $\tau$ in intra-layer coupling of one or both the layers in multiplex network can give rise to new features not being witnessed in the absence of delay. The dynamics of multiplexed layers with inclusion of time delay in one layer would be given by 
\mathleft
\begin{align}
\dot{\theta}_{1}^i = \omega^i_{1} +  {\lambda} \sum_{j=1}^N A^{ij}_{1} \sin(\theta^j_{1}- \theta^i_{1}) 
 + D \sin(\theta^i_{2}-\theta^i_{1}), \nonumber\\
 \setlength{\mathindent}{0pt}
m\ddot{\theta}^i_2 + \dot{\theta}^i_{2} = \omega^i_{2} + {\lambda} \sum_{j=1}^N A^{ij}_{2} \sin(\theta^j_{2}(t-\tau)-\theta^i_{2}(t)) \nonumber \\
 + D \sin(\theta^i_{1}-\theta^i_{2}).\nonumber
\end{align}\label{model3} 
\mathcenter
Delays have shown to control nature of phase transition~\citep{Ajaydeep2019} and relay synchronization in multiplex networks~\citep{Sawicki2018}. These results naturally form a background to extend 
the multiplex model system considered here to investigate role of delay on emerging ES. The interesting concern would be if phase lag and delay impart the same effect, and if the scenario of one
layer with delay and another layer without delay is favorable or detrimental for ES.

\subsection{A more general multiplexing strategy}
This article has focused on one-to-one coupling between the multiplexed layers, whereas real-world complex systems may have more complex multiplexing structure~\cite{Aleta2019}.  Many biological systems can be better modeled using one to many connections between the layers, for example neural networks immersed in the Glial cell medium (Fig.~\ref{fig:figure8}) \citep{Makovkin2017}. Instead of one to one coupling between the mirror nodes of different layers, one can define a
general bipartite network connecting nodes of the layer one to those of the layer two.

\subsection{Different dynamical evolution of individual node}
Instead of considering Kuramoto oscillators with and without inertia, one can consider other models such as FitzHugh-Nagumo (FHN) oscillators~\citep{FitzHugh1961} which is basically a model for neuronal networks. Hence, the dynamics of two multiplexed layers instead can be modeled by FHN oscillators in their oscillatory and excitable regimes. 
The occurrence of chimera state in two layered brain network model has been shown using FHN neuronal dynamics~\citep{Ling2019}. It is an interesting open problem to investigate possibilities of occurrence of ES for individual layer of multilayer networks governed by FHN oscillators in different regimes.

\section*{Acknowledgments}
\noindent SJ acknowledges Government of India, CSIR grant 25(0293)/18/EMR-II, DST grant EMR/2016/001921, and BRNS grant 37(3)/14/11/2018-BRNS/37131 for financial support. ADK acknowledges Govt. of India, CSIR grant 25(0293)/18/EMR-II for RA-ship. HJ acknowledges the National Research Foundation of Korea (NRF-2017R1A2B3006930) and KAIX program.

%% References with bibTeX database:
\bibliographystyle{elsarticle-num}

\bibliography{jcos_biblio}

\end{document}